\documentclass{easychair}

\DeclareGraphicsExtensions{.pdf,.jpeg,.png}

\usepackage{doc}
\usepackage{makeidx}
\usepackage[boxed]{algorithm2e}
\usepackage{color}
\usepackage{subcaption}
\usepackage{multirow}

\begin{document}

\title{Simulating Congestion Dynamics of Train \\Rapid Transit using Smart Card Data
\thanks{We thank the Land Transport Authority of Singapore for providing the anonymized smart card data. We thank Dr. Gary Lee Kee Khoon and Dr. Terence Hung Gih Guang for discussions. This research is supported by the Science and Engineering Research Council of the Agency for Science, Technology and Research (A*STAR) of Singapore (Complex Systems Programme grant number 122 45 04056), and also the A*STAR Computational Resource Centre through the use of its high performance computing facilities.}}

\titlerunning{Simulating Congestion Dynamics of Train Rapid Transit using Smart Card Data}

\author{
Nasri Bin Othman
  \and
Erika Fille Legara
	\and
Vicknesh Selvam
	\\
  \and
Christopher Monterola\thanks{Corresponding author}
}

\institute{
	Complex Systems Programme, Institute of High Performance Computing,\\
	Agency for Science, Technology and Research,\\
	1 Fusionpolis Way, \#16-16, Connexis North, Singapore 138632\\
	\email{\{othmannb, legaraeft, selvamv, monterolac\}@ihpc.a-star.edu.sg}
}

\authorrunning{Nasri et al.}

\clearpage

\maketitle

\begin{abstract}
Investigating congestion in train rapid transit systems (RTS) in today's urban cities is a challenge compounded by limited data availability and difficulties in model validation. Here, we integrate information from travel smart card data, a mathematical model of route choice, and a full-scale agent-based model of the Singapore RTS to provide a more comprehensive understanding of the congestion dynamics than can be obtained through analytical modelling alone. Our model is empirically validated, and allows for close inspection of the dynamics including station crowdedness, average travel duration, and frequency of missed trains---all highly pertinent factors in service quality. Using current data, the crowdedness in all 121 stations appears to be distributed log-normally. In our preliminary scenarios, we investigate the effect of population growth on service quality. We find that the current population (2 million) lies below a critical point; and increasing it beyond a factor of $\sim10\%$ leads to an exponential deterioration in service quality. We also predict that incentivizing commuters to avoid the most congested hours can bring modest improvements to the service quality provided the population remains under the critical point. Finally, our model can be used to generate simulated data for analytical modelling when such data are not empirically available, as is often the case.

\end{abstract}

\section{Introduction}
To tackle rising population density in urban cities, transportation planners often construct train rapid transit systems (RTS) as a first step. Yet continued population growth forces the RTS to evolve towards increased complexity with more train lines added to satisfy demand. With the increased complexity, planners are confronted with the difficulty of predicting commuter ridership, route choices, and also the various outcomes of the system during disruptions. Moreover, increased station and train crowdedness in RTS lead to congestion, commuter discomfort, trip delays, and lowered overall service quality standards. It is therefore imperative that modern transportation models be capable of investigating not just the issues of efficient, robust and scalable transportation, but also of commuter comfort and satisfaction.

The introduction of smart card ticketing in RTS has serendipitously enabled large-scale data analytics into commuter travel behaviour~\cite{Bagchi2005464, Pelletier2011557}. Analytical and regression models have been developed to estimate commuters' spatio-temporal density~\cite{sun2012using}, identification of boarded trains~\cite{Kusakabe2010}, travel patterns~\cite{Chakirov2011}, and transit use variability~\cite{Morency2007193}. Yet, it is noted that the information captured by smart cards has limitations~\cite{Pelletier2011557}; for example, most datasets do not contain routing information as they capture information only at the entry and egress points of journeys.



In contrast to analytical and regression models, agent-based models (ABM) strive to model each individual agent in a manner most natural to the system at hand~\cite{Bonabeau14052002}. Essentially, an agent is autonomous and formulates decisions and interacts with other agents directly. By directly replicating the mechanics of the system, an ABM permits the observation of emergent phenomena that arise from the interactions of the agents with each other~\cite{Bonabeau14052002}---provided the mechanics are correctly characterised and the model is well-calibrated.

ABM has seen recent success in modelling large-scale transportation~\cite{Erath2012, neumann2011micro, Wahba2011}. However, there are not many studies which incorporate smart card data to drive RTS demand for better calibration. In our previous work~\cite{legara2013cascadeabstract}, we had leveraged upon anonymized travel smart card transactional data to synthesise travel demand for a smaller-scale agent-based model of the Singapore transit system involving only one of the operational train lines, and achieved a very close match between the simulated and empirical travel duration distributions. In that work, we also investigated the dynamics of the smaller-scale system with regard to population growth.


Here, we extend our previous work~\cite{legara2013cascadeabstract} by: 1)~expanding the model to cover all seven operational lines; 2)~adding a route-choice mechanism inferred statistically from empirical travel duration distributions~\cite{monterola2013routeabstract}; 3)~incorporating station-specific walk-times; 4)~investigating dynamics that were not directly measurable in our dataset, such as station crowdedness; and 5)~running further population growth scenarios. We validate our model by ensuring the travel duration distributions generated from our simulations are well-calibrated to the empirical dataset. This would lend strength to any inferences derived from our scenarios. Apart from these goals, ultimately, we strive to construct a simulation platform that can be used to evaluate the efficacy of proposed strategies in tackling current and future urban transportation issues.


\section{Data}
Our model is dependent on data for the following purposes: 1)~to construct the transit infrastructure, 2)~to instantiate the commuter agents corresponding to the actual travel demand, 3)~to calibrate the travel time components of the network, and 4)~to accurately model the commuters' decision making (e.g., route choice). 

Our main data source is the anonymized travel smart card dataset for public transport users in Singapore, obtained from the Land Transport Agency (LTA) of Singapore. This amounts to over 14 million train journey records for 2 million unique card IDs taken across a full week. A trip begins with a tap in of the smart card at the origin station, and terminates with a tap out at the destination station. Here, we use the following fields for each record: \emph{origin~(tap-in station)}, \emph{destination~(tap-out station)}, \emph{tap-in time}, and \emph{trip duration}. From the \emph{origin}, \emph{destination}, and \emph{tap-in time} fields, we can reconstruct the travel demand for any given origin-destination~(O-D) pair and time. The \emph{trip duration} field is used for validating the simulation.

To construct the train network in our model, we consulted publicly-available resources, including the LTA website\footnote{\url{http://www.publictransport.sg/content/publictransport/en/homepage/trainmap.html}\\Last accessed: January 2014}. This yields the set of all train stations, their connectivity, and travel time between two adjacent stations. Information regarding the first and last trains at each station are also publicly available, and is used to estimate the train dispatch schedule.

Lastly, we estimated the walking times along the stations in order to account for commuter locomotion. The measurements made are coarse and not empirically verified; however, they are sufficient since walking is usually the smallest component of travel---typically less than 2 minutes (shorter than the time taken for a train to travel between most adjacent stations).

\section{Methodology}
\begin{figure}
	\centering
		\includegraphics[width=1.0\textwidth]{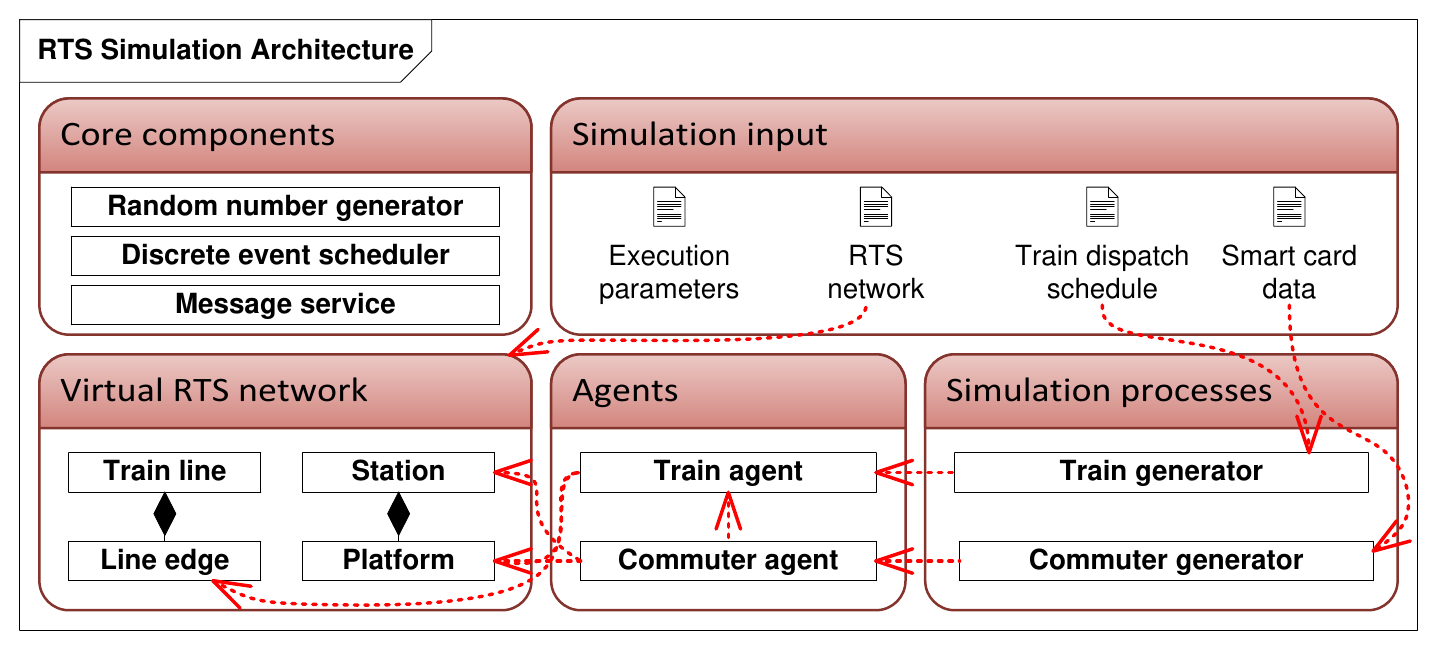}
	\caption{RTS simulation architecture. Components are grouped under five major categories. The dashed arrows indicate the influence/interaction of components to other components. The solid diamonds under the Virtual RTS network group indicate composition.}
	\label{fig:Architecture}
\end{figure}


Figure~\ref{fig:Architecture} shows the architecture of our agent-based model. As with many computational agent-based models, it is powered by a discrete event scheduler to schedule agents and processes; a random number generator to introduce stochasticity; and a message service to enable agent communication and interaction. Simulation input includes execution parameters, the RTS network, an estimated train dispatch schedule, and the smart card data comprising the journeys for the day being simulated. Two types of agents are modelled: trains and commuters.

The input RTS network---comprising 121 stations, 412 directed edges connecting adjacent stations, and 7 train lines---is used to construct a virtual network consisting of station entities connected via the given edges and lines. For each line connecting to a station, two platforms are created---one for each direction---where train agents may park and commuter agents may wait, board and alight trains. Stations connected to multiple lines are termed \emph{interchanges}. At any time, each station holds a collection (possibly empty) of commuter agents whose size determines the station \emph{crowdedness}. Commuter agents are inserted into the collection when they tap-in or alight from a train, and are removed when they tap-out or board a train. 

Our network data include estimations of walking time as it accounts for a statistically significant portion of travel duration. Walking time estimations are provided for every tapping gate-to-platform and platform-to-platform (for transfers) combinations. Commuter agents will have to walk from point to point in the station; for instance, they cannot immediately board trains upon tapping in but must first walk to the target platform. The actual walking time for each commuter agent is normally distributed around this estimation to introduce stochasticity.

We estimated the train dispatch schedule from information obtained in the LTA website. This schedule is used by the train generator process to construct and insert train agents into the simulation. The inter-event time between the insertion of trains on the same track is governed by the train dispatch frequency. Currently this is set to: 1.5--4.5 minutes for peak hours (7:30--9:30am and 5:30--7:30pm) and 4.5--7.5 minutes for off-peak hours---normally distributed (with truncation). Due to stochasticity, the total number of trains dispatched in a day is not fixed.

Train agents are inserted into their initial platforms, and will travel the span of the line until they have reached the final platforms. A train agent waits for a random amount of time in each platform (55--65 seconds if on an interchange; 30--40 seconds otherwise) before proceeding to the next platform. Commuter agents may only board and alight during the time the train agent waits on a platform. It will be removed shortly after it reaches the final platform. A train agent has a limited capacity for commuters (1920 for mass-rapid transit lines, 105 for light-rapid transit lines), and those who are unable to board although they have already reached the platform will have to wait for the next train and are considered to have missed the train.

We utilize the anonymized smart card data from LTA to construct the commuter agents. From each relevant record in the dataset, the commuter generator process constructs a commuter agent based on the following fields: 1)~tap-in time, 2)~origin station, and 3)~destination station. The commuter agent is inserted into the origin station at the specified tap-in time and will have to navigate the RTS to reach the destination station, where it can be removed by tapping-out. In order to reach its destination, the commuter agent may undertake the following sequence of actions: 1)~walking from tap-in gate to platform; 2)~waiting for a train; 3)~boarding a train; 4)~alighting a train; 5)~walking to transfer to a different platform (if necessary); and 6)~walking to the tap-out gate and tapping-out. The total time the commuter spends in the simulation from tapping in to tapping out determines the travel duration of the commuter. We emphasize that our simulation is not provided with the original journey's empirical travel duration; this duration is to be computed by our simulation. We can then compare the statistical distribution of travel duration from our simulation with the empirical data for validation.

As our full-scale model incorporates multiple RTS lines, there may be multiple routes for a commuter to take from origin (O) to destination (D). This is empirically evidenced by the multi-modality of the travel duration distributions for certain O-D pairs. We use the approach in~\cite{monterola2013routeabstract} to determine the routes for a given O-D as follows. First, the set of candidate routes are computed under certain constraints (e.g., cannot be much longer than shortest path or involve too many transfers). Then, the empirical probability density function is fitted using one or more Gumbel distribution components. Each component is matched against a candidate route by comparing the mean duration and the path distance of the route. The integral of the component determines the probability that the route is chosen. From the set of route-probability pairs, we can then probabilistically assign a route for each commuter agent.

\section{Evaluation of agent-based model}

\subsection{Validation}
We validate our model by comparing the distributions of travel duration derived from our simulations against the empirical data, for each day in our dataset (one week). To reduce standard error, only origin-destination~(O-D) pairs with a demand of at least 2000 are compared. For each O-D pair, we collected the travel duration observations from 30 simulation runs and the empirical sample (for that day), and computed the trip duration density functions via kernel density estimation. 
Goodness-of-fit is then computed through the following measures: 1)~Bhattacharyya's coefficient~\cite{bhattacharyya1943measure, Chung1989280}~($BC$), which measures the amount of overlap in the distributions; 2)~probability plot correlation coefficient~\cite{filliben1975probability}~($PPCC$), which measures similarity up to linearity; and 3)~Linfoot's criteria~\cite{Huck:85}, which compare the \emph{shape} of the density curves and are comprised of Fidelity~($F$), Structural Content~($C$), and Correlation Quantity~($Q$). Their formulations are given in Table~\ref{tab:SimilarityMeasures}.

\begin{figure}
	\centering
	\begin{subfigure}{.4\textwidth}
		\includegraphics[width=\textwidth]{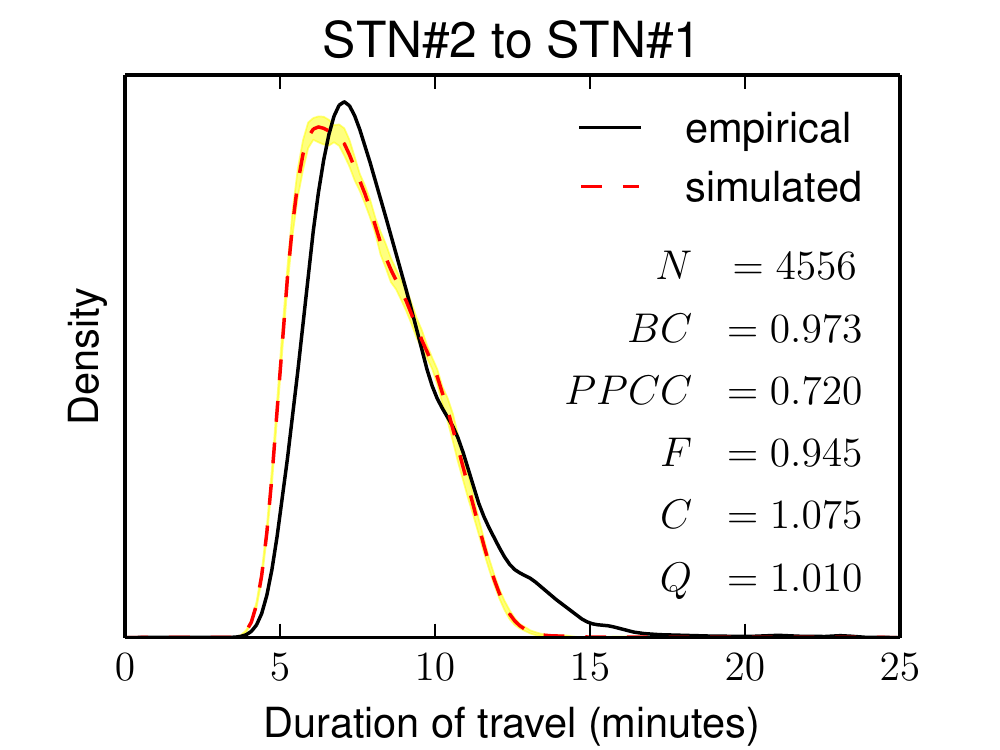}
		\caption{Most frequented O-D}
		\label{fig:most_populated_od}
	\end{subfigure}
	\begin{subfigure}{.4\textwidth}
		\includegraphics[width=\textwidth]{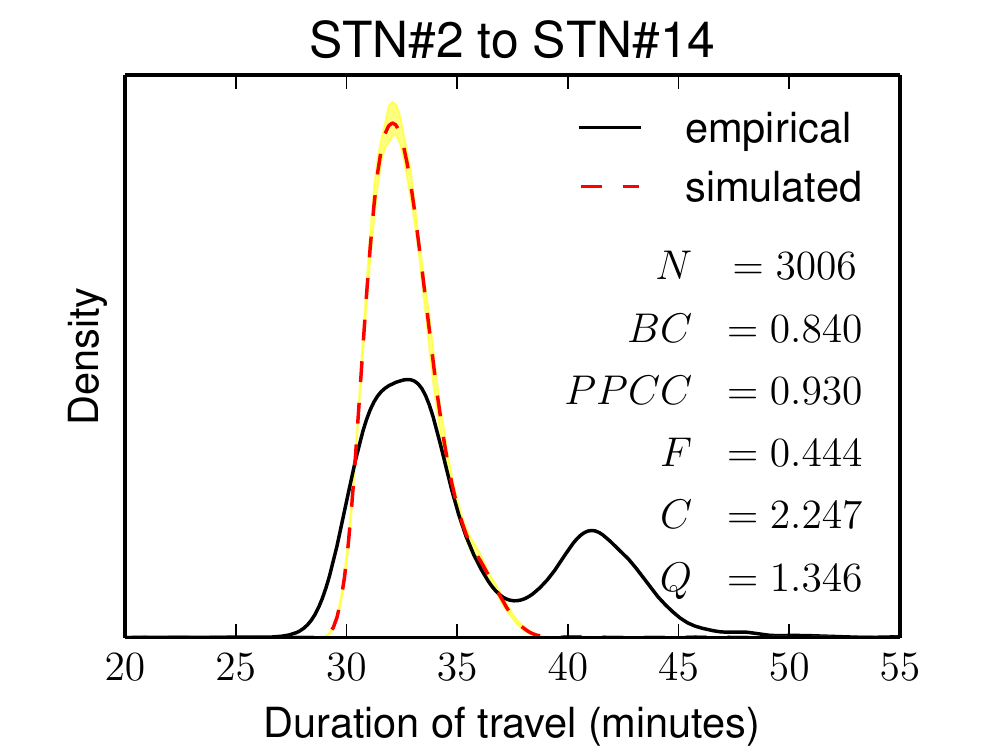}
		\caption{Penultimate station effect}
		\label{fig:penultimate_station_effect}
	\end{subfigure}
	\caption[Trip duration density for simulation and empirical]{Trip duration density for simulation (dashed) and empirical (solid). Yellow band indicates the inter-quartile range from 30 simulation samples. (\subref{fig:penultimate_station_effect}) shows the penultimate station effect, manifested as a secondary peak in the empirical curve but not the simulated curve. }
	\label{fig:od}
\end{figure}

The goodness-of fit results, averaged over 30 runs, are shown in Table~\ref{tab:SimulationGoodnessOfFit}. See also Figure~\ref{fig:od} for goodness of fit for two selected O-D pairs. With $BC$ above 0.9, our results are in good agreement with the empirical data across the entire week. The $PPCC$ and $F$ measures are lower (0.7-0.8). This is possibly due to their formulations which are highly sensitive to noise in the simulation and empirical samples. Note that the observed empirical durations compose a \emph{single sample} for the demand in a given O-D and day; the \emph{true distribution} is unknown.




\begin{table}
	\centering
		\begin{tabular}{lllcl}
		\textbf{Measure} & & \textbf{Symbol} & \textbf{Computation}\\ \hline \hline \noalign{\vskip 2mm} 
		\multicolumn{2}{l}{Bhattacharyya's coefficient~\cite{bhattacharyya1943measure, Chung1989280}} & $BC$ & $ \int \sqrt{p(t)q(t)} dt$\\ \noalign{\vskip 2mm} \hline \noalign{\vskip 2mm}
		\multicolumn{2}{l}{Probability plot correlation coefficient~\cite{filliben1975probability}} & $PPCC$ & $\frac{\sum{(X_i - \bar{X})(Y_i - \bar{Y})}}{\sigma_X\sigma_Y}$ \\ \noalign{\vskip 2mm} \hline \noalign{\vskip 2mm}
		Linfoot's criteria~\cite{Huck:85}& Fidelity & $F$ & $1-\frac{\int{(q(t)-p(t))^2}dt}{\int{p(t)^2}dt}$ \\ \noalign{\vskip 2mm}  \cline{2-4} \noalign{\vskip 2mm}
		& Structural Content & $C$ & $\frac{\int{q(t)^2}dt}{\int{p(t)^2}dt}$ \\ \noalign{\vskip 2mm}  \cline{2-4} \noalign{\vskip 2mm}
		& Correlation Quantity & $Q$ & $\frac{\int{p(t) q(t)}dt}{\int{p(t)^2}dt}$ \\ \noalign{\vskip 2mm} \hline \noalign{\vskip 2mm}
		\multicolumn{4}{l}{$p(t)$--Empirical duration density; $q(t)$--Simulation duration density;} \\
		\multicolumn{4}{l}{$X$--Sorted empirical duration observations; $Y$--Sorted simulation duration observations} \\ 
		\end{tabular}
	\caption{Goodness-of-fit measures}
	\label{tab:SimilarityMeasures}
\end{table}

\begin{table}
	\centering
		\begin{tabular}{lcccccc}
			& & & & \multicolumn{3}{c}{Linfoot's criteria} \\ 
			Day & O-D pairs & $\overline{BC}$	& $\overline{PPCC}$	& $\overline{F}$	& $\overline{C}$	& $\overline{Q}$ \\ \hline \hline
			
			Monday	& 89	& 0.940	& 0.819	& 0.813	& 1.138	& 0.975 \\
			Tuesday	& 97	& 0.934	& 0.823	& 0.767	& 1.167	& 0.967 \\
			Wednesday	& 102	& 0.932	& 0.841	& 0.780	& 1.223	& 1.001 \\
			Thursday	& 100	& 0.934	& 0.841	& 0.781	& 1.245	& 1.013 \\
			Friday	& 121	& 0.941	& 0.821	& 0.818	& 1.139	& 0.978 \\
			Saturday	& 92	& 0.940	& 0.763	& 0.840	& 1.086	& 0.963 \\
			Sunday	& 52	& 0.926	& 0.737	& 0.813	& 1.310	& 1.061 \\ \hline \hline
			Overall	& 653	& 0.936	& 0.812	& 0.801	& 1.178	& 0.990	
		\end{tabular}
	\caption{Simulation goodness-of-fit for all O-D pairs above 2000 demand, averaged over 30 runs}
	\label{tab:SimulationGoodnessOfFit}
\end{table}

\subsection{Observed dynamics}
\begin{figure}
	\centering
	\begin{subfigure}{.4\textwidth}
		\includegraphics[width=\textwidth]{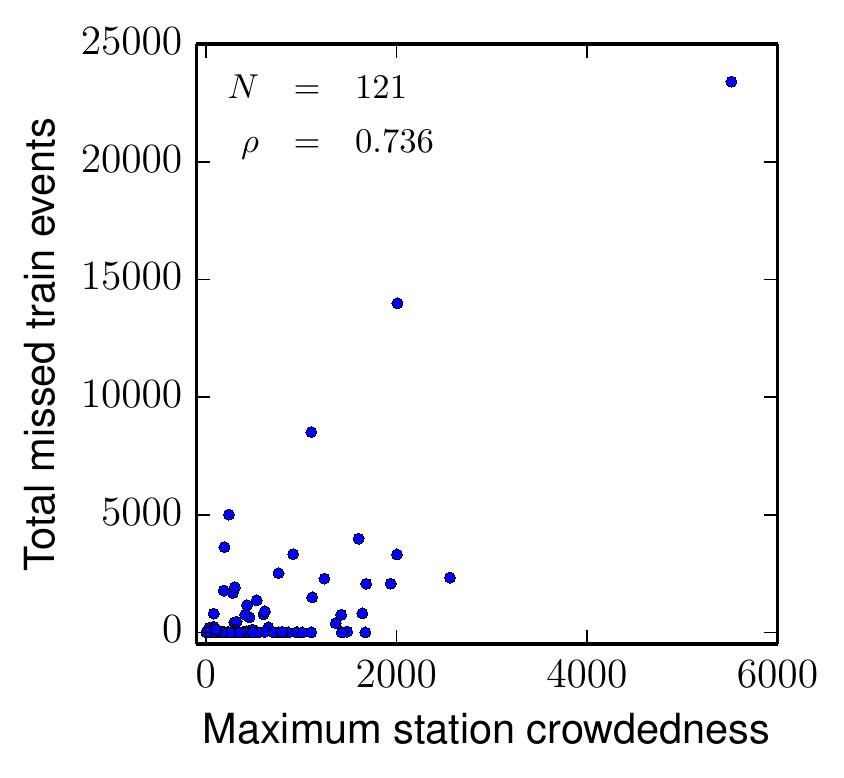}
		\caption{Correlation}
		\label{fig:plot_crowdedness_and_missed_trains_a}
	\end{subfigure}
	\begin{subfigure}{.4\textwidth}
		\includegraphics[width=\textwidth]{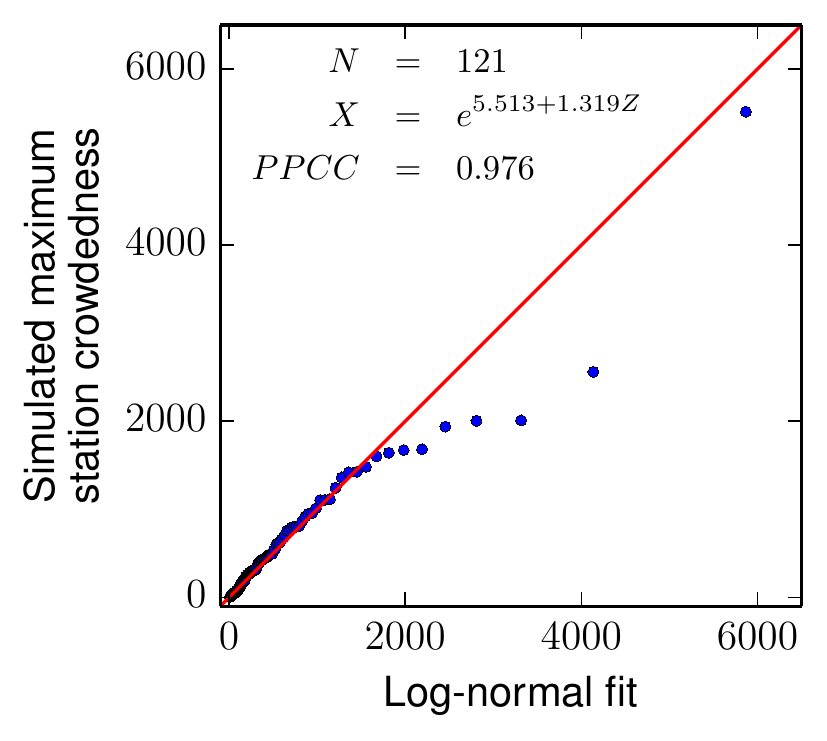}
		\caption{Quantile-quantile plot}
		\label{fig:plot_crowdedness_and_missed_trains_b}
	\end{subfigure}
	\caption{Simulated maximum station crowdedness and total missed train events experienced in 121 RTS stations throughout Monday. (a)~Correlation between each station's max crowdedness and missed trains. Most stations, even with high crowdedness, do not appear to have significant missed train events. A weak positive correlation is seen (Pearson's correlation, $\rho=0.736$). (b)~Quantile-quantile plot comparing the simulated maximum crowdedness and a log-normal fit. The most crowded stations appear to be overestimated by the fit.}
	\label{fig:plot_crowdedness_and_missed_trains}
\end{figure}

A key strength of our agent-based model is that it can be used to measure system phenomena that are not directly available from data, or easily inferred using an analytical (i.e., closed-form) model. In Figure~\ref{fig:plot_crowdedness_and_missed_trains}, we show the computed measurements of maximum station crowdedness and missed trains. Most stations do not experience significant missed train events even when crowded, leading to a weak positive correlation. Maximum crowdedness and missed train events are typically highest in the interchanges and their surrounding stations (not depicted). Interestingly, we find that maximum crowdedness appears to be log-normally distributed according to the following equation:

\[X=\exp(5.513 + 1.319Z),\]

 where X is maximum crowdedness of a station and Z is a standard-normal variable, with a probability plot correlation coefficient~($PPCC$)~\cite{filliben1975probability} of 0.976 for the 121 stations. No such fit holds for missed train events. These measurements are highly sensitive to the input train dispatch schedule, and further calibration is required before these results can be meaningfully interpreted. Once calibrated properly, it would be interesting to utilise these measurements in analytical RTS models as a substitute for empirical data, which are often unavailable.

Chakirov~\cite{Chakirov2011} inferred that commuters boarding from penultimate (second-to-last) stations tend to travel backwards to the last station and re-board in order to attain a seat for longer journeys even at the cost of additional travelling time. The empirical distribution in Figure~\ref{fig:penultimate_station_effect} shows the effect of this \emph{penultimate station effect} on the travel duration, manifesting as a secondary peak. As our agent-based model does not account for this effect, the secondary peak does not appear in the simulated distribution. Yet, since our model captures the supply-side (i.e., train) dynamics accurately, we may conclude that the secondary peak is indeed caused \emph{solely} by the commuters' behaviour; rather than overcrowding or missed trains. This is reinforced by the observation that our simulation still matches well for trips starting from the same station but to nearer destinations, whereby the \emph{penultimate station effect} is less pronounced. An extension of our model will be required to take this effect into account.



\section{Preliminary scenarios}
\subsection{Scenario descriptions}
A major concern to urban planners is the scalability of their transit systems with regard to population growth. In our preliminary scenarios, we adjust the transit demand in our agent-based model to predict how population growth may affect commuter experiences with respect to travel duration and the number of trains missed. Monte Carlo sampling is performed on the 2,078,010 journeys in the Monday dataset to generate the desired transit demand. Scaling using the Monte Carlo method implies that the generated demand tends to be proportional to the original demand. Scaling beyond the original demand would introduce duplicated journeys, but this is acceptable since the duplicated commuters are likely to have different walking speeds. In addition to scaling the population, we introduce a station crowdedness limit, $\Psi$, which when exceeded will disallow new commuters from tapping in. This limit is tripled for interchanges~($3\Psi$). This mechanism is used to prevent impossible levels of crowdedness as well as to provide a measure of scalability by counting the rejected commuters. Additionally, we expanded the peak hour train dispatch timings to 6--11am and 4--9pm, as it is the least that can be done by operators to alleviate peak congestion. Two preliminary scenarios are investigated: (A)~adjusting the crowdedness limit; and (B)~re-shaping of commuter demand.

In our previous work~\cite{legara2013cascadeabstract}, a smaller scale model revealed the existence of a critical point whereby scaling the commuter population further causes exponential increases in travel duration. Here, in Scenario A, we scale the population from 78,010 to 6,078,010 in steps of 100,000, using the following values for station crowdedness limit, $\Psi \in \{\infty, 9000, 7000, 5000, 3000\}$. The purpose of this scenario is to replicate the results in our previous work, and also to determine if the station crowdedness limit (previously unaccounted) would affect the dynamics observed.

Currie~\cite{currie2010quick} highlighted that a program in Australia to incentivize commuters to complete their morning journeys earlier resulted in 23\% participation and hence reduced the travel demand during the morning peak. With this in mind, in Scenario B, we scale the population from 2,078,010 (original) to 6,078,010 in steps of 100,000, fix $\Psi = 3000$, and temporally reshape the commuter demand during the most congested hours to investigate whether the reshaped demand will bring improvements. A participation ratio, $\Phi~\in~\{0.00, 0.05, 0.10, 0.20, 0.30\}$, determines the proportion of commuters originally tapping-in at the most congested hours who will adjust their tap-in times. Participating commuters who would have tapped-in at 7--9am would instead tap-in at 6--7am; and those who tapped in at 6--8pm would instead tap-in at 8--9pm. 40.2\% of journeys in the empirical data have tap-ins within those original ranges.

\subsection{Scenario results}
\begin{figure}[t]
	\centering
		\includegraphics[width=1.0\textwidth]{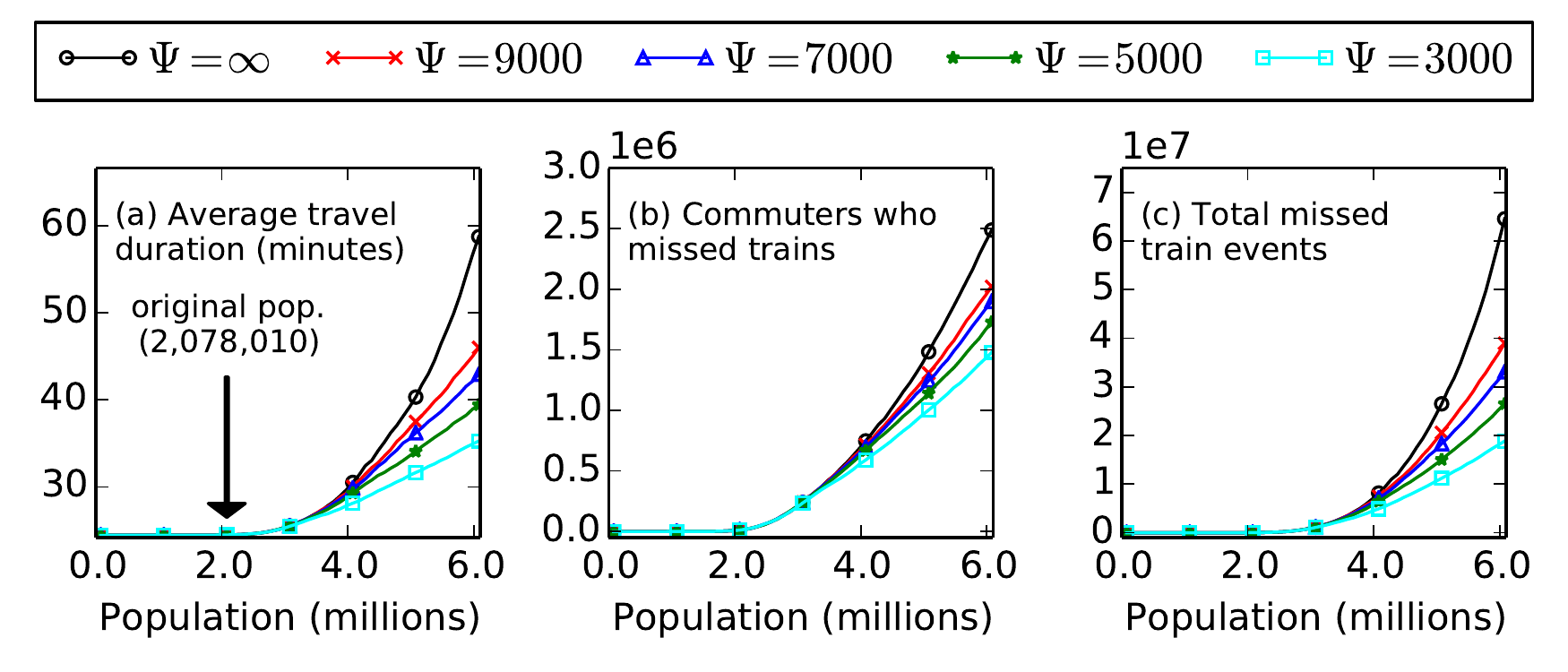}
	\caption{Population scaling results for scenario A showing the relationship between scaling population (in steps of 100K) and: (a)~average travel duration, (b)~number of commuters who missed trains, and (c)~number of missed train events---with respect to the station crowdedness limit, $\Psi$. Markers are plotted every 500K population. (a) and (c) are \emph{not} linear transformations of each other. This is so even if we divide (c) by the current population to get an average.}
	\label{fig:plot_adjust_population}
\end{figure}
\begin{figure}[ht]
	\centering
		\includegraphics[width=1.0\textwidth]{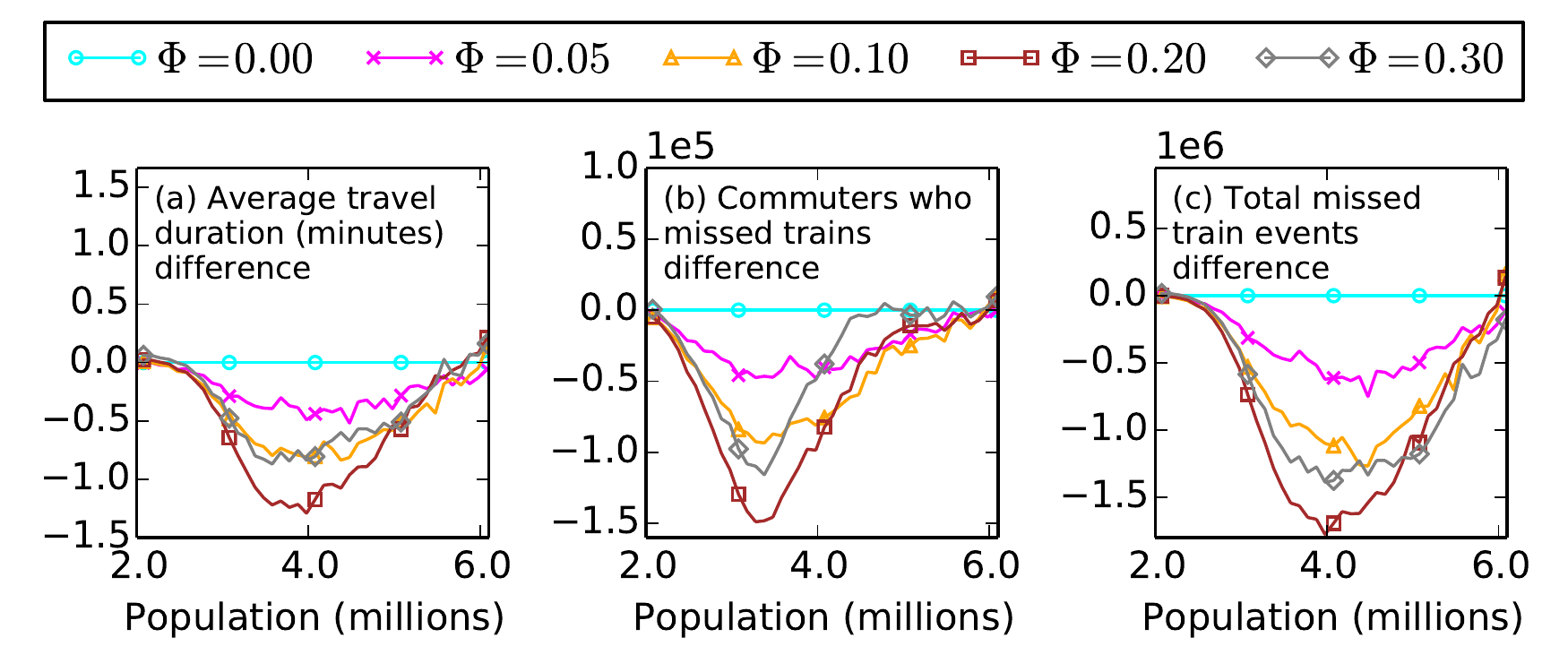}
	\caption{Population scaling (with reshaping) results for scenario B showing the \emph{differences} between reshaping ($\Phi>0$) and no reshaping ($\Phi=0.00$, horizontal line). For reference, the actual values for the $\Phi=0.00$ case are shown in Figure~\ref{fig:plot_adjust_population} with $\Psi=3000$. Reshaping yields improvement in almost all cases, with $\Phi=0.20$ displaying the best improvements.}
	\label{fig:plot_adjust_population_reshaping}
\end{figure}

Figure~\ref{fig:plot_adjust_population} shows the results for Scenario A. A critical point is seen around the 2--3M population range, whereby all three measures begin exhibiting exponential growth. The location of the point appears to be invariant with regard to $\Psi$, but is different for each indicator. Decreasing $\Psi$ leads to significant reduction in all three indicators. Nonetheless, even with $\Psi=3000$, the critical point still exists. At 6M population, over 652K of commuters were rejected under the $\Psi=3000$ case. This experiment shows that commuters who tap-in when the station is already overcrowded will unduly impact the service quality measures (significantly more than other commuters), and should thus be rejected and provided with alternative means of transportation.

Figure~\ref{fig:plot_adjust_population_reshaping} shows the results for Scenario B. Noting that the differences are much smaller in magnitude compared to the original curves in Figure~\ref{fig:plot_adjust_population}, it is clear that reshaping yields only a modest improvement at best. Nonetheless, in almost all cases, it is clearly advantageous for commuters to avoid the most congested hours. 20\% participation of commuters provides the best improvement to service quality. This is coincidentally similar to the 23\% participation ratio reported by Currie~\cite{currie2010quick}.

All in all, these results show that the current RTS infrastructure in Singapore is running close to its critical capacity. A \emph{proportional increase} in the population by $\sim10\%$ can likely be endured with marginal increases to travel duration, but the frequency of missed trains experienced will be nearly tripled ($22$K to $62$K). Beyond the critical point, even measures such as expanding peak hours and reshaping demand temporally still prove insufficient, suggesting that more comprehensive strategies must be used to tackle congestion (e.g., increasing train frequency; adding new lines; reshaping demand \emph{spatio-temporally} and not just temporally).

\section{Conclusion}
In this work, we had incorporated empirically-derived travel demand data into a full-scale agent-based model of the train rapid transit system in Singapore. Our approach granted us a more comprehensive view of the congestion dynamics than afforded by analysing the anonymized smart card dataset directly. By modelling every commuter individually, we were able to synthesise highly detailed measurements, including crowdedness and number of trains missed. With these measurements, transport operators can more accurately estimate the comfort and satisfaction of commuters. They may then construct strategies to maximise comfort and satisfaction and relieve congestion, in addition to the traditional objectives of efficient transport. 


The usefulness of an agent-based transport simulation goes beyond being able to accurately estimate the current RTS dynamics. In our preliminary scenarios, we demonstrated how our agent-based model was used to predict the outcomes of various strategies under population growth (e.g., expanded peak hours, crowdedness limits, temporally reshaped demand). Such predictions are highly invaluable to transportation planners, but would be difficult to achieve using analytical and regression models alone. Since agent-based models require good calibration to make accurate and reliable predictions, it is critical that further calibration be performed against more empirical data, including when the RTS experiences exceptional conditions such as disruptions and special occasions (e.g., New Year's Eve).

One implication of our work is that it should be possible to apply this approach to RTS deployments in other countries, so long as demand (empirical or synthetic), train dispatch schedules, walk times, and network structure are available. Indeed, our results matched well with the empirical travel durations using input largely derived from publicly available information---with the exception of the anonymized smart card dataset and estimated walk times.


Finally, train rapid transit represents only one means of travel for commuters, albeit often it metaphorically represents the backbone of transport in an urban city. Other means of travel available in Singapore include public bus services, taxis, private vehicles, and walking. It is common for one person in one week to be involved in several of these depending on the type, time and distance of journey and the socio-economic status of the person. Indeed, these transport modes are inter-dependent, and in our future work, we will be looking into integrating several transport models together to capture a holistic and dynamic view of transport in Singapore.


\phantomsection
\addcontentsline{toc}{section}{References}
\bibliographystyle{plain}
\bibliography{ref}

\begin{thebibliography}{10}

\bibitem{Bagchi2005464}
M.~Bagchi and P.R. White.
\newblock The potential of public transport smart card data.
\newblock {\em Transport Policy}, 12(5):464 -- 474, 2005.

\bibitem{bhattacharyya1943measure}
Anil Bhattacharyya.
\newblock On a measure of divergence between two statistical populations
  defined by their probability distributions.
\newblock {\em Bull. Calcutta Math. Soc}, 35(99-109):4, 1943.

\bibitem{Bonabeau14052002}
Eric Bonabeau.
\newblock Agent-based modeling: Methods and techniques for simulating human
  systems.
\newblock {\em Proceedings of the National Academy of Sciences}, 99(suppl
  3):7280--7287, 2002.

\bibitem{Chakirov2011}
Artem Chakirov and Alexander Erath.
\newblock Use of public transport smart card fare payment data for travel
  behaviour analysis in {Singapore}.
\newblock In {\em 16th International Conference of Hong Kong Society for
  Transportation Studies}, December 2011.

\bibitem{Chung1989280}
J.K. Chung, P.L. Kannappan, C.T. Ng, and P.K. Sahoo.
\newblock Measures of distance between probability distributions.
\newblock {\em Journal of Mathematical Analysis and Applications}, 138(1):280
  -- 292, 1989.

\bibitem{currie2010quick}
Graham Currie.
\newblock Quick and effective solution to rail overcrowding.
\newblock {\em Transportation Research Record: Journal of the Transportation
  Research Board}, 2146(1):35--42, 2010.

\bibitem{Erath2012}
A.~Erath, P.~Fourie, M.~Eggermond, S.~Ordóñez, A.~Chakirov, and K.~Axhausen.
\newblock Large-scale agent-based transport demand model for {Singapore}.
\newblock In {\em 13th International Conference on Travel Behaviour Research
  (IATBR)}, July 2012.

\bibitem{filliben1975probability}
James~J. Filliben.
\newblock The probability plot correlation coefficient test for normality.
\newblock {\em Technometrics}, 17(1):111--117, 1975.

\bibitem{Huck:85}
F.O. Huck, C.L. Fales, N.~Halyo, R.W. Samms, and K.~Stacy.
\newblock Image gathering and processing: information and fidelity.
\newblock {\em J. Opt. Soc. Am. A}, 2(10):1644--1666, October 1985.

\bibitem{Kusakabe2010}
Takahiko Kusakabe, Takamasa Iryo, and Yasuo Asakura.
\newblock Estimation method for railway passengers' train choice behavior with
  smart card transaction data.
\newblock {\em Transportation}, 37(5):731--749, 2010.

\bibitem{legara2013cascadeabstract}
E.F. Legara, C.~Monterola, T.~Hung, and G.~Lee.
\newblock Cascade of travel duration delays in a rapid transport system
  [abstract].
\newblock In {\em European Conference on Complex Systems Book of Abstracts},
  pages 94--95. Complex Systems Society and complexitat.CAT, September 2013.

\bibitem{monterola2013routeabstract}
C.~Monterola, E.F. Legara, P.~Di, G.~Lee, and T.~Hung.
\newblock Non-invasive procedure to probe the route choices of commuters in
  rail transit systems [abstract].
\newblock In {\em European Conference on Complex Systems Book of Abstracts}.
  Complex Systems Society and complexitat.CAT, September 2013.

\bibitem{Morency2007193}
Catherine Morency, Martin Trépanier, and Bruno Agard.
\newblock Measuring transit use variability with smart-card data.
\newblock {\em Transport Policy}, 14(3):193 -- 203, 2007.

\bibitem{neumann2011micro}
A.~Neumann and M.~Balmer.
\newblock Micro meets macro: A combined approach for a large-scale,
  agent-based, multi-modal and dynamic transport model for {Berlin}.
\newblock Technical report, VSP Working Paper 11-14, TU Berlin, Transport
  Systems Planning and Transport Telematics, 2011., 2011.

\bibitem{Pelletier2011557}
M.P. Pelletier, M.~Trépanier, and C.~Morency.
\newblock Smart card data use in public transit: A literature review.
\newblock {\em Transportation Research Part C: Emerging Technologies},
  19(4):557 -- 568, 2011.

\bibitem{sun2012using}
L.~Sun, D.H. Lee, A.~Erath, and X.~Huang.
\newblock Using smart card data to extract passenger's spatio-temporal density
  and train's trajectory of {MRT} system.
\newblock In {\em Proceedings of the ACM SIGKDD International Workshop on Urban
  Computing}, pages 142--148. ACM, 2012.

\bibitem{Wahba2011}
Mohamed Wahba and Amer Shalaby.
\newblock Large-scale application of {MILATRAS}: case study of the {Toronto}
  transit network.
\newblock {\em Transportation}, 38(6):889--908, 2011.

\end{thebibliography}

\label{sect:bib}

%
%
%
%
%

\end{document}